\documentstyle[12pt]{article}
\newcommand{\beq}{\begin{equation}}
\newcommand{\eeq}{\end{equation}}
\newcommand{\ba}{\begin{eqnarray}}
\newcommand{\ea}{\end{eqnarray}}

\newcommand{\dsl}
  {\kern.06em\hbox{\raise.15ex\hbox{$/$}\kern-.56em\hbox{$\partial$}}}

\newcommand{\eeqarr}{\end{eqnarray}}
\newcommand{\ZZ}{{\rm \kern 0.275em Z \kern -0.92em Z}\;}
\begin{document}
\begin{titlepage}
\begin{center}
\vspace*{0.5cm}

{\Huge A Note on Scalar Field Theory}\\ \vspace*{0.5cm}{\Huge in
$AdS_{3}/CFT_{2}$}\\ \vspace*{1.5cm} {\large Pablo
Minces}\\
\vspace*{0.5cm} Instituto de Astronom\'{\i}a y F\'{\i}sica del
Espacio
(IAFE),\\
C.C.67 - Suc. 28, 1428 Buenos Aires, Argentina.\\
minces@iafe.uba.ar
\end{center}
\vspace*{1.5cm}
\begin{abstract}
We consider a scalar field theory in $AdS_{d+1}$, and introduce a
formalism on surfaces at equal values of the radial coordinate. In
particular, we define the corresponding conjugate momentum. We
compute the Noether currents for isometries in the bulk, and perform
the asymptotic limit on the corresponding charges. We then introduce
Poisson brackets at the border, and show that the asymptotic values
of the bulk scalar field and the conjugate momentum transform as
conformal fields of scaling dimensions $\Delta_{-}$ and
$\Delta_{+}$, respectively, where $\Delta_{\pm}$ are the standard
parameters giving the asymptotic behavior of the scalar field in
AdS. Then we consider the case $d=2$, where we obtain two copies of
the Virasoro algebra, with vanishing central charge at the classical
level. An $AdS_{3}/CFT_{2}$ prescription, giving the commutators of
the boundary CFT in terms of the Poisson brackets at the border,
arises in a natural way. We find that the boundary CFT is similar to
a generalized ghost system. We introduce two different ground
states, and then compute the normal ordering constants and quantum
central charges, which depend on the mass of the scalar field and
the AdS radius. We discuss certain implications of the results.
\end{abstract}

\end{titlepage}
\section{Introduction}
An intensive study of diverse theoretical aspects of Anti-de Sitter
(AdS) spaces has been carried out since the proposal
\cite{maldacena} of the existence of a duality between a
supergravity theory on AdS and a Conformal Field Theory (CFT) living
at its boundary. In addition, the precise AdS/CFT prescription given
in \cite{witten}\cite{gubser}, where the partition function of the
AdS theory is identified with the generating functional of the dual
CFT, has allowed to perform several explicit checks and
calculations.

In this context, the scalar field theory on AdS space is an
interesting toy model which allows to analyze diverse aspects of the
AdS/CFT correspondence and exhibits some subtle properties, so that
it has received considerable attention in the literature. For
instance, the early works \cite{freedman}\cite{freedman2} (see also
\cite{mezincescu}) showed that it possesses the interesting property
of having two different kinds of normalizable modes, thus giving
rise to two possible quantizations in the AdS bulk. This happens for
masses of the scalar field in the range \ba m_{BF}^{2}\; <\;
m^{2}\;<\; m_{BF}^{2}\; +\;\frac{1}{l^{2}}\;\; , \label{4} \ea where
\ba m_{BF}^{2}\; =\; -\frac{d^{2}}{4l^{2}}\;\; , \nonumber \ea is
the Breitenlohner-Freedman mass. Here $d+1$ is the dimension of the
$AdS_{d+1}$ space and $l$ is the AdS radius. The
Breitenlohner-Freedman bound reads \ba m^{2}\;\geq\; m_{BF}^{2}\;\;
, \label{1771} \ea and solutions below it correspond to tachyons in
AdS. Throughout this note, we will consider masses of the scalar
field in the range (\ref{4}).\footnote{We will exclude the
particular case $m^{2}\; =\;m_{BF}^{2}$ from our analysis.}

We will propose here a new approach to the formulation of scalar
field theory in the AdS/CFT correspondence, which we hope allows to
gain further insight in the way both theories relate to each other,
and will lead to find interesting new results. Even when some
calculations will be performed for AdS spaces of generic $d+1$
dimensions, our main focus here will be on the $d=2$ case.

In order to set our notation, we point out that throughout this note
we will consider the Euclidean representation of $AdS_{d+1}$ in
Poincar\'e coordinates, described by the half space $x^{0}>0$,
$x^{i} \in {\bf R}$ with metric \ba ds^{2}=\frac{l^{2}}{x_{0}^{2}}
\sum_{\mu=0}^{d} dx^{\mu}dx^{\mu}\;\; . \label{1} \ea In particular,
the boundary of the AdS space is located at $x^{0}\rightarrow 0$.

The behavior of the scalar field close to the border is of the
form\footnote{We point out that the quantization in
\cite{freedman}\cite{freedman2} was performed in global coordinates.
We take the results which are relevant to our present purposes.} \ba
\Phi (\epsilon,\vec{x})\; =\; \epsilon^{\Delta_{+}}(\alpha
(\vec{x})+ O(\epsilon^{2}))\; +\; \epsilon^{\Delta_{-}}(\beta
(\vec{x})+ O(\epsilon^{2}))\;\; ,\label{7} \ea where $\epsilon\; =\;
x^{0}$ is taken to be small. Here we have \ba \Delta_{\pm}\; =\;
\frac{d}{2}\; \pm\; l\sqrt{m^{2}\; -\; m_{BF}^{2}}\;\; . \label{23}
\ea The usual procedure is to require boundary conditions that set
to zero half of the modes of the field near the boundary, thus
giving rise to two possible quantizations.

From the AdS/CFT point of view, we expect to find two different CFTs
at the border. One of them, corresponding to a dual operator of
conformal dimension $\Delta_{+}$, was reproduced through the
prescription in \cite{witten} (see also \cite{freedman3}\cite{mueck}
for further analysis). The other one, corresponding to the conformal
dimension $\Delta_{-}$, was found to be obtained by performing a
Legendre transformation to the original generating functional in the
theory with conformal dimension $\Delta_{+}$ \cite{kw} (see also
\cite{kraus} for previous results). Related issues involve
double-trace perturbations and the role of boundary conditions for
scalar field theory in Anti-de Sitter space (see e.g.
\cite{witten2}-\cite{andrade}).

Following the standard AdS/CFT prescription in
\cite{witten}\cite{gubser}, most of the literature dealing with the
scalar field theory in the AdS/CFT correspondence focuses on
correlation functions. As expected, this has proven a fruitful
approach. However, in this note we would like to consider the
particular case of three dimensional AdS space and shift the focus
to the information we could get about the boundary two dimensional
CFT by considering the `conserved' charges of the theory in the
bulk. The motivation is as follows. We know that isometries of the
$AdS_{3}$ background correspond to global conformal transformations
at the boundary. Taking this into account, our proposal here is to
identify the generators of global conformal transformations of the
boundary CFT with the asymptotic expressions of the charges in the
bulk, and then perform a proper expansion on such generators in
order to compute the Virasoro generators of the theory. This idea is
somehow similar in spirit to that analyzed in \cite{bh} (see also
e.g. \cite{henneaux}\cite{henneaux2}\cite{hollands} for the
inclusion of the scalar field theory into the analysis), but we will
consider it here in the context of a formulation where we make use
of the fixed metric (\ref{1}) and choose $x^{0}$ to play a special
role, so that we foliate the space on surfaces at equal values of
it, as we will discuss shortly.

Now, once in possession of the expressions of the Virasoro
generators of the boundary CFT, we aim at proposing a proper
prescription which allows to compute the Virasoro algebra including
the corresponding central charge, which is in principle expected to
depend on the mass of the scalar field and the AdS radius, in a
similar way as the conformal dimensions of the boundary CFT
operators do through the usual AdS/CFT prescription in
\cite{witten}\cite{gubser}. The computation of the Virasoro algebra
in the asymptotic limit would constitute a non trivial result which
could be considered as a consistency check on our calculations. In
addition, to compute the corresponding central charge would provide
some interesting new information on the boundary CFT and on the way
both theories in different dimensions relate to each other. Besides,
it is expected that some other information on the boundary CFT could
be obtained, e.g. from the requirement for the central charge to be
positive, or from the specific form the Virasoro generators would
have in terms of the asymptotic expressions of quantities in the
bulk. At a more speculative level, possible applications e.g.
statistical or in black hole physics could also exist, but we will
not address this issue here.

Now, the computation of the generators of the boundary CFT or the
calculation of their corresponding algebra could in principle be
performed only after developing a formalism where all the
information in the bulk is mapped to the boundary. In particular,
this concerns the above mentioned fact that, for masses in the range
(\ref{4}), there are two possible quantizations of the bulk scalar
field. We should be able to take this information to the boundary.
In order to do this, there is the important observation to be made
that not only the bulk field, but also the corresponding canonical
momentum in a formulation where the radial coordinate $x^{0}$ (see
(\ref{1})) plays the role of `time', should be taken into account. A
formalism which makes use of this canonical momentum was presented
in \cite{skenderis}\cite{papadi}. The fact that its inclusion is
required follows from results in \cite{mr}\cite{m1}\cite{m2} (a
discussion of other aspects of this issue is postponed to footnote
10 since it will better be considered after the introduction of some
results and notation). The point is that the inclusion of both the
field and the momentum is required in order to have the complete
information about both boundary CFTs corresponding to both
quantizations in the bulk.

Motivated by the discussion above, we will then consider the radial
coordinate $x^{0}$ to play a special role, and introduce a
formulation on surfaces at equal values of it. In particular, we
will define the corresponding conjugate momentum. In this way, when
carefully performing the limit $\epsilon\rightarrow 0$ we will
locate on the surface $x^{0}=0$, where Poisson brackets of
`conserved' charges may be computed. Such charges will be obtained
from the Noether currents corresponding to isometries on the $AdS$
bulk, and we will show them to be finite in the limit
$\epsilon\rightarrow 0$, provided the action is supplemented by a
proper surface term.

In order to perform certain consistency checks and introduce aspects
of the formalism, in Section 2 we will first focus in the generic
case of $AdS_{d+1}$. On the one hand, we will verify that the
asymptotic charges actually generate the global conformal algebra in
$d$ dimensions. From well known results in the literature we also
expect, in addition, that the asymptotic values of the bulk scalar
field and the momentum, which we will call $\Phi_{0}$ and $\Pi_{0}$,
should transform as conformal fields of scaling dimensions
$\Delta_{-}$ and $\Delta_{+}$, respectively. Using the Poisson
brackets, we will verify that the asymptotic charges actually
realize these required properties too. Even when expected, these
results are non trivial, and any of them can be considered as a
consistency check on our formalism.

We point out that, before the checks above are performed, we will
have to consider the interesting and important issue of the well
definiteness of the limit $\epsilon\rightarrow 0$ when it is
performed on the charges. We will show this to hold when the action
for the scalar field is supplemented by precisely the same (already
known in the literature) boundary term which makes the action to be
finite in such limit, thus adding to the consistency of the
formalism.

Having performed all the checks above, we will then in Section 3
focus on the case of $AdS_{3}$, where the main results of this note
will be obtained. Here the boundary will be described in terms of
complex holomorphic and antiholomorphic coordinates, and the
definition of the charges will involve contour integrals. Once
again, we will perform the limit $\epsilon\rightarrow 0$, and then,
by Laurent expanding the asymptotic charges we will obtain two
series of coefficients $L_{n}$ and ${\bar L}_{n}$ $(n\in {\bf Z})$.
On the one hand, we will show these coefficients to satisfy the
expected but non trivial result of giving rise to two copies of the
Virasoro algebra, with vanishing central charge at the classical
level. On the other hand, we will show $\Phi_{0}$ and $\Pi_{0}$ to
transform as conformal fields of weights $(\frac{\Delta_{-}}{2} ,
\frac{\Delta_{-}}{2})$ and
$(\frac{\Delta_{+}}{2},\frac{\Delta_{+}}{2})$, respectively, in
agreement with the previous results in Section 2. Motivated by this,
we will also propose mode expansions for $\Phi_{0}$ and $\Pi_{0}$,
and reproduce again the same results as before, this time in terms
of modes. Once again, any of the results above can also be
considered as a non trivial check on the formalism.

At this point, we will put aside the role of $\Phi_{0}$ and
$\Pi_{0}$ as the asymptotic values of the bulk fields, and treat
them as conformal fields with the given weights, living in the
boundary CFT. We will then consider aspects of the quantization of
such CFT. The motivation to attempt this is that, as we will see, a
prescription relating the Poisson brackets at $x^{0}=0$ to
commutators on the boundary CFT will arise in a natural way. This
will be a non trivial prescription, since it will relate Poisson
brackets in the asymptotic limit of a three dimensional theory to
commutators in a two dimensional one. This will be allowed by the
property of the asymptotic Poisson brackets of being computed at
equal values of the distance to the origin of the complex plane (in
a manner to be illustrated later), as well as, simultaneously, at
the surface of fixed $x^{0}=0$. This property will give the
asymptotic Poisson brackets a meaning from the point of view of the
boundary CFT, and will be inherited from the definition of the
charges using contour integrals. It is also what will allow us to go
one step further in Section 3 than in the generic $AdS_{d+1}$ case
of Section 2, where we will deal with a $d+1$ dimensional theory
only.

These calculations will lead us to find expressions for the
generators and commutators of the theory which, exception made of
the fact that the fields will not factorize in the holomorphic and
antiholomorphic parts, are surprisingly similar to the corresponding
ones in generalized ghost systems.

Then, we will introduce two different ground states, which are not
$SL_{2}$ invariant, and that correspond to choosing the zero mode of
which one of the fields, $\Phi_{0}$ or $\Pi_{0}$, is grouped with
the lowering operators. This should correspond to the two possible
quantizations in the bulk found in \cite{freedman}\cite{freedman2}.
The fact that we will find two different quantizations, as expected,
can be considered as a last non trivial check on our formalism.

In both cases we will find the same normal ordering constants and
quantum central charges, which are given by $L_{0}|0\rangle={\bar
L}_{0}|0\rangle=\frac{l^{2}m^{2}}{8}|0\rangle$ and $c={\bar
c}=2+3l^{2}m^{2}$, respectively. As expected, they depend on the
mass of the bulk field and the AdS radius, in an analogous way as
the conformal dimensions do. Thus, our formalism allows to find
interesting new information on the boundary CFT corresponding to
scalar field on $AdS_{3}$, and on the way both theories in different
dimensions relate to each other. We point out that, as we will show,
the normal ordering constant and central charge are in the ranges
$-\frac{1}{8}<\frac{l^{2}m^{2}}{8}<0$ and $-1<c<2$, respectively. In
particular, we will show that the requirement for the central charge
to be positive sets $l^{2}m^{2}>-\frac{2}{3}$. This condition is
more restrictive than the Breitenlohner-Freedman bound. It is a new
and interesting result, which seems to be detected only from the
boundary point of view. The precise meaning of this and the way in
which it could be red from the bulk point of view deserve more
studies.

A surprising result that we will also find will be that, from the
fact that the boundary CFT will be similar to a generalized ghost
system, and from the explicit dependence of the central charge with
the mass of the scalar field, we will conclude that the mass of the
bulk scalar field seems to play the role of a background charge for
the boundary CFT. We believe that further insight into this
interesting issue could be obtained by performing a bosonization
program on the boundary CFT. However, we will no longer pursue this
issue here.

\section{Asymptotic limit in $AdS_{d+1}$}

The action of a massive, minimally coupled scalar field theory in
$AdS_{d+1}$ is of the form \ba I_{0}=\frac{1}{2}\;\int_{{\cal M}}
d^{d+1} x \;\sqrt{g}\; \left(
g^{\mu\nu}\partial_{\mu}\Phi\;\partial_{\nu}\Phi\; + \;
m^{2}\Phi^{2}\right) \;\; , \label{2} \ea where $m$ is the mass of
the scalar field and $g_{\mu\nu}$ is given by (\ref{1}). Our
conventions are that Greek indices $\mu , \nu ,...$ take the values
from 0 to d. The equation of motion reads $(\nabla^{2}-m^{2})\Phi
=0$.

We consider the space as foliated by a family of surfaces $\partial
M_{\epsilon}$ defined by $x^{0}=\epsilon$, and with outward pointing
unit normal vector \ba n_{\mu}=(-l\epsilon^{-1},{\bf 0})\; .
\nonumber \ea In particular, the boundary $\partial M$ of the AdS
space is located at $\epsilon\rightarrow 0$. Actually, the action
includes terms which diverge in such limit, and in order to take
care of them we supplement $I_{0}$ with a proper surface term, which
will not introduce any changes in the equation of motion. We take
the action to be \ba I\; =\; I_{0}\; +\; \sigma \; I_{S}\; ,
\label{10} \ea where $\sigma$ is a coefficient and \ba
I_{S}=\int_{\cal M} d^{d+1} x
\;\sqrt{g}\;\nabla_{\mu}(n^{\mu}\Phi^{2})\;\; . \label{74} \ea In
fact, the choice of $\sigma$ for which there are no divergent terms
in the action, for $\epsilon\rightarrow 0$, and provided we consider
the range (\ref{4}), is given by \cite{m1}\cite{mr}\footnote{See
also \cite{m2}\cite{marolf1}\cite{marolf2}. In
\cite{skenderis}\cite{papadi} Eq.(\ref{16}) has been derived using
the conjugate momentum in a formulation on surfaces at equal values
of $x^{0}$. Here we will focus on the well definiteness of the limit
$\epsilon\rightarrow 0$ when it is performed on the charges.
Actually (\ref{16}) works well in the asymptotic limit of both the
charges and the action, as expected, and as already suggested by the
results in \cite{mr}.} \ba \sigma =\frac{\Delta_{-}}{2l}\;\; .
\label{16} \ea However, we will maintain a generic value of $\sigma$
for a while, because it will be interesting to see how (\ref{16})
will arise again in a different context.

We consider isometries of the AdS background, i.e. coordinate
transformations $x^{\mu}\rightarrow x^{\mu}+\delta x^{\mu}$ such as
$\delta g_{\mu\nu}=0$. The variation of the action (\ref{10}) is of
the form \ba \delta I\sim \int_{\partial{\cal M}_{\epsilon}}
d^{d}x\; \sqrt{h}\; n_{\mu}\; J^{\mu}\; , \nonumber \ea where
$h_{\mu\nu}$ is the induced metric and $J^{\mu}$ is the Noether
current.\footnote{The standard approach would be to consider the
following definition of the conserved current ${\hat J}^{\mu}\sim
T^{\mu}_{\;\;\nu}\delta x^{\nu}$, where $T_{\mu\nu}$ is the usual
energy-momentum tensor of the scalar field obtained by performing an
infinitesimal variation on the metric, $g_{\mu\nu}\rightarrow
g_{\mu\nu}\; +\; \delta g_{\mu\nu}$, and writing \ba \delta_{g}I\;
=\; \frac{1}{2}\int d^{d+1}x\;\sqrt{g}\; T_{\mu\nu}\;\delta
g^{\mu\nu}\;\; .\nonumber\ea As emphasized in \cite{mr}, $J^{\mu}$
and ${\hat J}^{\mu}$ are inequivalent. The reasons why we make use
of $J^{\mu}$ instead of ${\hat J}^{\mu}$ are, on the one hand, that,
given that we want to compute the generators of the asymptotic
symmetries corresponding to isometries in the bulk, then it seems
the natural choice to make, and, on the other hand, the arguments
presented in detail in \cite{mr}. In particular, ${\hat J}^{\mu}$
does not contain the information on (\ref{16}), and this happens so
because it is unsensitive to the addition of the surface term
(\ref{74}) to the action (see \cite{mr} for additional details).}
The calculations are similar to those in \cite{mr}\footnote{Actually
only the Noether current for time displacements in global
coordinates was computed in \cite{mr}, but at this stage the formal
calculations are similar, even when we are considering a different
metric and generic isometries.} and we find \ba J^{\mu}\; =\;
\Lambda^{\mu}_{\;\;\nu}\;\delta x^{\nu}\; , \label{11} \ea where \ba
\Lambda_{\mu\nu} &=& \Theta_{\mu\nu}\; +\; \sigma\;
[g_{\mu\nu}\nabla_{\alpha}(n^{\alpha}\Phi^{2})\; -\; 2\; n_{\mu}\Phi
\;\partial_{\nu}\Phi]\;\; ,
\nonumber\\
\Theta_{\mu\nu} &=& -\partial_{\mu}\Phi\; \partial_{\nu}\Phi\;+\;
\frac{1}{2}\; g_{\mu\nu}(g^{\alpha\beta}\partial_{\alpha}\Phi\;
\partial_{\beta}\Phi\; +\; m^{2}\Phi^{2})\;\; .
\label{11a}
\ea

We write \ba I\;=\;\int_{\cal M} d^{d+1}x\; \sqrt{g}\; {\cal L}\;\;
, \nonumber \ea where the Lagrangian is given by \ba {\cal L}\; =\;
\frac{1}{2}\;
\left(g^{\mu\nu}\partial_{\mu}\Phi\;\partial_{\nu}\Phi\; + \;
m^{2}\Phi^{2}\right) \; +\;\sigma\;\nabla_{\mu}(n^{\mu}\Phi^{2})\;\;
. \nonumber \ea As discussed above, an important point regarding our
calculations is that the usual equal time formalism will be replaced
here by a formulation on surfaces at equal values of the radial
coordinate $x^{0}$.\footnote{This will eventually allow us to
consider the boundary at $x^{0}=0$ as a particular choice among such
surfaces, and compute Poisson brackets on it.} We introduce the
conjugate momentum \ba \Pi\; =\; \sqrt{g}\; \frac{\partial {\cal
L}}{\partial (\partial_{n}\Phi)}\; =\; \sqrt{g}\; (\partial_{n}\Phi
\; + \; 2\;\sigma \;\Phi)\;\; , \label{8} \ea where
$\partial_{n}\Phi= n^{\mu}\partial_{\mu}\Phi =
-\frac{x^{0}}{l}\partial_{0}\Phi$ is the normal
derivative.\footnote{Eq.(\ref{8}) can also be written in the
following form \ba\Pi\; = \; -\sqrt{h}\;\left(\partial_{0}\Phi\; -\;
\frac{2\;l\;\sigma}{x^{0}}\;\Phi\right)\;\; ,\nonumber\ea which,
exception made of the second term in the r.h.s., is closer to the
usual notation. Here $h_{\mu\nu}$ is the induced metric.} So we can
write \ba \Lambda^{0}_{\;\; i} &=&
\frac{\Pi}{\sqrt{g}}\;\frac{x^{0}}{l}\;\partial_{i}\Phi\;\; ,
\nonumber\\
\Lambda^{0}_{\;\; 0} &=& 2\sigma\; \frac{\Pi}{\sqrt{g}}\; \Phi\; -\;
\frac{\Pi^{2}}{2g}\; + \; \frac{1}{2}\;\partial^{i}\Phi\;
\partial_{i}\Phi\; +\; \frac{1}{2l}\; (lm^{2}\; +\;
2\sigma d\; -\; 4\sigma^{2}l)\; \Phi^{2}\; , \label{20} \ea where we
have adopted the convention that Latin indices i,j,... take the
values from 1 to d.

So far, we have considered generic isometries of the AdS background.
In order to determine the Noether currents (\ref{11}) we need to
explicitly write $\delta x^{\mu}$ in terms of Killing vectors. These
are given by \cite{frolov} \ba \xi^{0}_{T} &=& 0\qquad ,\qquad
\xi^{i}_{T}\;\; =\;\; a^{i}\;\;
;\nonumber\\
\xi^{0}_{R} &=& 0\qquad , \qquad \xi^{i}_{R}\;\; =\;\; m^{i}_{\;  j}x^{j}
\qquad (m_{ji}\; =\; -m_{ij})\;\; ;\nonumber\\
\xi^{\mu}_{D} &=& \alpha x^{\mu}\quad ;\nonumber\\
\xi^{0}_{S} &=& 2x^{0}x^{i}b^{i}\;\; , \;\;\;\xi^{i}_{S} \;\; =\;\;
2x^{i}x^{j}b^{j}-x^{j}x^{j}b^{i}-x^{0}x^{0}b^{i}\;\; . \label{12}
\ea Here $\xi^{\mu}_{T}$, $\xi^{\mu}_{R}$, $\xi^{\mu}_{D}$ and
$\xi^{\mu}_{S}$ act at the boundary as translations, rotations,
dilations and special conformal transformations, respectively. This
suggests to consider that the charges \ba Q=\int_{\partial{\cal
M}_{\epsilon}} d^{d}x\; \sqrt{g}\; J^{0}\; , \label{13} \ea act at
the boundary on $\epsilon\rightarrow 0$ as generators of conformal
transformations.

Using (\ref{11}, \ref{12}, \ref{13}) we find the following charges
\ba P_{i} &=&\int_{\partial{\cal M}_{\epsilon}} d^{d}x\;
\sqrt{g}\;\Lambda^{0}_{\;\; i}\qquad ,\qquad M_{i}^{\; j}\;\;
=\;\;\int_{\partial{\cal M}_{\epsilon}} d^{d}x\;
\sqrt{g}\;(\Lambda^{0}_{\;\;
i}\; x^{j}\; -\; \Lambda^{0j}\;x_{i})\;\; ,\nonumber\\
\nonumber\\
D &=&\int_{\partial{\cal M}_{\epsilon}} d^{d}x\;
\sqrt{g}\;\Lambda^{0}_{\;\; \mu}\; x^{\mu}\; ,\qquad K^{i}\;\;
=\;\;\int_{\partial{\cal M}_{\epsilon}} d^{d}x\; \sqrt{g}\;(2\;
\Lambda^{0}_{\;\; \mu}\; x^{\mu}\; x^{i}\; -\;
\Lambda^{0i}\;x_{\mu}\; x^{\mu})\; . \nonumber\\ \label{15} \ea

The next step is to compute the charges above in the limit
$\epsilon\rightarrow 0$. We assume that the asymptotic behavior of
the scalar field is given by the lowest order term in (\ref{7}),
namely \ba \Phi(\epsilon,
\vec{x})\;\sim\;\epsilon^{\Delta_{-}}\;\Phi_{0}(\vec{x})\; .
\label{14} \ea Then from (\ref{8}) the conjugate momentum should
approach the boundary as \ba \Pi(\epsilon, \vec{x})\;\sim\;
\left(2\;\sigma\;
-\;\frac{\Delta_{-}}{l}\right)\;\epsilon^{-\Delta_{+}-1}\;\Phi_{0}(\vec{x})\;
. \label{9} \ea Now, when plugging (\ref{14}, \ref{9}) into
(\ref{15}) we see that in general the charges diverge in the limit
$\epsilon\rightarrow 0$. However, we notice from (\ref{9}) that when
$\sigma$ is chosen as in (\ref{16}) the
$O(\epsilon^{-\Delta_{+}-1})$ term does not contribute. In such
case, the asymptotic behavior of $\Pi$ should be obtained by
plugging the next lowest order term of (\ref{7}) into (\ref{8}).
Such next order term of (\ref{7}) has to be chosen among the
possibilities $O(\epsilon^{\Delta_{-}+\; 2})$ and
$O(\epsilon^{\Delta_{+}})$. Since we are considering the range
(\ref{4}) then the lowest order of the two is
$O(\epsilon^{\Delta_{+}})$, and using (\ref{8}) this gives \ba
\Pi(\epsilon,
\vec{x})\;\sim\;\epsilon^{-\Delta_{-}-1}\;l\;\Pi_{0}(\vec{x})\; ,
\label{18} \ea which replaces (\ref{9}). Notice that (\ref{16})
simplifies (\ref{20}) to \ba \Lambda^{0}_{\;\; i} &=&
\frac{\Pi}{\sqrt{g}}\;\frac{x^{0}}{l}\;\partial_{i}\Phi\;\; ,
\nonumber\\
\Lambda^{0}_{\;\; 0} &=& \frac{\Delta_{-}}{l}\;
\frac{\Pi}{\sqrt{g}}\; \Phi\; -\; \frac{\Pi^{2}}{2g}\; + \;
\frac{1}{2}\;\partial^{i}\Phi\;\partial_{i}\Phi\;\; , \label{21} \ea
and the conjugate momentum (\ref{8}) now reads \ba \Pi\; =\;
\sqrt{g}\;\left(\partial_{n}\Phi \; + \; \frac{\Delta_{-}}{l}
\;\Phi\right)\; . \label{22} \ea Plugging (\ref{14}, \ref{18},
\ref{21}) into (\ref{15}) and taking into account (\ref{4}) we see
that the charges are finite in the limit $\epsilon\rightarrow
0$.\footnote{In particular, notice that only the first term in the
r.h.s of the expression of $\Lambda^{0}_{\;\; 0}$ (see (\ref{21}))
will contribute, since the other two terms are of higher order in
$\epsilon$. In addition, $\Lambda^{0}_{\;\; i}$ has a non-vanishing
contribution.} Thus, from now on we choose $\sigma$ to be given by
(\ref{16}) and the expressions (\ref{21}, \ref{22}), together with
the asymptotic behaviors (\ref{14}, \ref{18}), are
assumed.\footnote{The choice (\ref{16}) was found in
\cite{mr}\cite{m1}\cite{m2} to correspond to the case where the
divergent local terms in the asymptotic expression of the action
vanish, and, in addition, to make the canonical energy computed in
global coordinates to be conserved, positive and finite for
`irregular' modes propagating in the bulk. Here the choice
(\ref{16}) arises again, now making the charges to be finite in the
asymptotic limit.}

A comment is in order. Notice that the asymptotic behaviors of $\Pi$
and $\Phi$ were obtained using the first and second contributions to
the r.h.s. of (\ref{7}), respectively. This suggests that, in the
asymptotic limit, the scaling dimension of $\Pi$ should be
$\Delta_{+}$, whereas $\Phi$ should have dimension $\Delta_{-}$.
This fact will explicitly be verified shortly, and will play an
important role in what follows.\footnote{The facts that $\Pi$
approaches the boundary as in (\ref{18}), and that it should
transform with scaling dimension $\Delta_{+}$ in the asymptotic
limit, were shown already in the calculations in
\cite{mr}\cite{m1}\cite{m2} (see also \cite{m3} for previous related
results). In particular, in the notation in \cite{m2}, it was shown
that, once (\ref{16}) is chosen, the field `$\psi^{(1)}$' given by
$\psi^{(1)}=\partial_{n}\Phi+\frac{\Delta_{-}}{l}\Phi$ (i.e. $\Pi
=\sqrt{g}\psi^{(1)}$, see (\ref{22})) approaches the boundary as
$\psi^{(1)}(\epsilon,
\vec{x})\sim\epsilon^{\Delta_{+}}\psi^{(1)}(\vec{x})$, from which
(\ref{18}) follows. In addition, it was shown that $\psi^{(1)}$
couples, through the standard AdS/CFT prescription given in
\cite{witten}\cite{gubser}\cite{kw}, with a modification explained
in \cite{mr}, to a boundary conformal operator of dimension
$\Delta_{-}$, thus indicating that $\psi^{(1)}$, and by extension
$\Pi$, should have scaling dimension $\Delta_{+}$ in the asymptotic
limit. On the other hand, the interpretation of $\Pi$ as the
conjugate momentum to $\Phi$ (see (\ref{8})) was given in
\cite{skenderis}\cite{papadi}.}

Now, taking the limit $\epsilon\rightarrow 0$ and using (\ref{4},
\ref{14}, \ref{18}, \ref{21}) we find the following expressions for
the charges (\ref{15}) evaluated on the boundary at $x^{0}=0$ \ba
{\tilde P}_{i} &=& -\int d^{d}x\;\Pi_{0}\;\partial_{i}\Phi_{0}\qquad
,\qquad {\tilde M}_{i}^{\; j}\;\; =\;\;-\int d^{d}x\; \Pi_{0}\;
(x^{j}\;\partial_{i}\; -\; x_{i}\;\partial^{j})\;\Phi_{0}\;\;
,\nonumber\\
{\tilde D} &=& -\int d^{d}x\; \Pi_{0}\; (x^{i}\;\partial_{i}\; + \;
\Delta_{-})\;\Phi_{0}
\;\; ,\nonumber\\
{\tilde K}^{i}  &=&  -\int d^{d}x\;\Pi_{0}\; [2\;x^{i}\;
(x^{j}\;\partial_{j}+\Delta_{-})\; -\;
x^{j}\;x_{j}\;\partial^{i}]\;\Phi_{0}\;\; , \label{24} \ea where
${\tilde P}_{i}\equiv \lim_{\epsilon\rightarrow 0}P_{i}$ and similar
definitions are assumed for the remaining charges.

As we have pointed out before, we are considering here a formulation
where $x^{0}$ plays a special role and we step on surfaces at equal
values of it. In particular, we have now positioned on the boundary
at $x^{0}=0$, where we consider the Poisson brackets \ba
{\{}\Pi_{0}(\vec{x}),\Phi_{0}(\vec{y}){\}}_{P.B.} \;=\;
\delta^{d}(\vec{x}-\vec{y})\;\; , \label{25}\ea and \ba
{\{}\Pi_{0}(\vec{x}),\Pi_{0}(\vec{y}){\}}_{P.B.}
\;=\;{\{}\Phi_{0}(\vec{x}),\Phi_{0}(\vec{y}){\}}_{P.B.} \;=\; 0\;\;
, \label{255} \ea which give \ba {\{}{\tilde P}_{i}\;
,\;\Phi_{0}{\}}_{P.B.} &=& -\;\partial_{i}\Phi_{0}\qquad ,\qquad
{\{}{\tilde M}_{i}^{\; j}\; ,\;\Phi_{0}{\}}_{P.B.} \;\;=\;\; -\;
(x^{j}\;\partial_{i}\; -\; x_{i}\;\partial^{j})\;\Phi_{0}\;\; ,
\nonumber\\
{\{}{\tilde D}\; ,\;\Phi_{0}{\}}_{P.B.} &=& -\;
(x^{i}\;\partial_{i}\; + \; \Delta_{-})\;\Phi_{0}\quad ,
\nonumber\\
{\{}{\tilde K}^{i}\; ,\;\Phi_{0}{\}}_{P.B.}&=& -\; [2\;x^{i}\;
(x^{j}\;\partial_{j}+\Delta_{-})\; -\;
x^{j}\;x_{j}\;\partial^{i}]\;\Phi_{0}\;\; . \label{75} \ea In this
way, the asymptotic charges ${\tilde P}_{i}$, ${\tilde M}_{i}^{\;
j}$, ${\tilde D}$ and ${\tilde K}^{i}$ respectively generate
translations, rotations, dilations and special conformal
transformations at the boundary. Notice, in addition, that
$\Phi_{0}$ has scaling dimension $\Delta_{-}$, as expected.

Now, performing integrations by parts in (\ref{24}) and using
(\ref{23}), we also find \ba {\{}{\tilde P}_{i}\;
,\;\Pi_{0}{\}}_{P.B.} &=& -\;\partial_{i}\Pi_{0}\qquad ,\qquad
{\{}{\tilde M}_{i}^{\; j}\; ,\;\Pi_{0}{\}}_{P.B.} \;\;=\;\; -\;
(x^{j}\;\partial_{i}\; -\; x_{i}\;\partial^{j})\;\Pi_{0}\;\; ,
\nonumber\\
{\{}{\tilde D}\; ,\;\Pi_{0}{\}}_{P.B.} &=& -\;
(x^{i}\;\partial_{i}\; + \; \Delta_{+})\;\Pi_{0}\quad ,
\nonumber\\
{\{}{\tilde K}^{i}\; ,\;\Pi_{0}{\}}_{P.B.}&=& -\; [2\;x^{i}\;
(x^{j}\;\partial_{j}+\Delta_{+})\; -\;
x^{j}\;x_{j}\;\partial^{i}]\;\Pi_{0}\;\; , \label{76} \ea thus
showing that $\Pi_{0}$ has scaling dimension $\Delta_{+}$, as
anticipated. In the following section, this result will motivate us
to treat $\Phi_{0}$ and $\Pi_{0}$ as independent fields with the
given dimensions.

Using (\ref{25}, \ref{255}), the asymptotic charges (\ref{24}) can
analogously be shown to satisfy the global conformal algebra in $d$
dimensions, as expected. This result, as well as (\ref{75}) and
(\ref{76}), can be considered as non trivial checks on our
formalism.

In this way, we have computed the explicit expressions of the
asymptotic charges which generate the conformal transformations at
the boundary. They are written in terms of the asymptotic values of
the bulk scalar field and the conjugate momentum, namely $\Phi_{0}$
and $\Pi_{0}$, which in turn behave as conformal fields with scaling
dimensions $\Delta_{-}$ and $\Delta_{+}$, respectively.

\section{Asymptotic limit in $AdS_{3}$}

We now turn our attention to the particular case of $d=2$, where we
expect to find the local conformal algebra and Virasoro generators
in the asymptotic limit. We also expect to compute the corresponding
central charge and find related information.

We consider the following change of variables \ba (x^{0}\; ,\;
x^{1}\; ,\; x^{2})\;\longrightarrow\; (x^{0}\; ,\; z\; ,\; {\bar
z})\;\; , \nonumber \ea where the complex variables $z$ and ${\bar
z}$ are given by \ba z\; =\; x^{1}\; +\; i\;x^{2}\qquad ,\qquad
{\bar z}\; =\; x^{1}\; -\; i\;x^{2}\;\; . \label{26} \ea The status
of $z$ and ${\bar z}$ is similar here to the usual one in standard
two dimensional CFT, in that we extend the range of $x^{1}$ and
$x^{2}$ to the complex plane, and so (\ref{26}) is understood as
just a change of independent variables. In this way, ${\bar z}$ is
not the complex conjugate of $z$, namely $z^{*}$. On the other hand,
the physical space is the `real' surface ${\bar z}=z^{*}$ where we
recover $x^{1},x^{2}\in {\bf R}$.

The non-vanishing components of the metric are \ba g_{00}\; =\;
\frac{l^{2}}{x_{0}^{2}}\qquad ,\qquad g_{z{\bar z}}\; =\; g_{{\bar
z}z}\; =\; \frac{l^{2}}{2x_{0}^{2}}\;\; , \label{27} \ea and the
Killing vectors (\ref{12}) now read \ba \xi^{0}_{T} &=& 0\quad
,\quad \xi^{z}_{T}\;\; =\;\; a^{1}\; +\; i\; a^{2} \quad ,\quad
\xi^{{\bar z}}_{T}\;\; =\;\; a^{1}\; -\; i\; a^{2}\;\; ; \nonumber\\
\xi^{0}_{R} &=& 0\quad ,\quad \xi^{z}_{R}\;\; =\;\; -\; i\;
m^{1}_{\; 2}\; z\quad ,\quad \xi^{{\bar z}}_{R}\;\; =\;\; i\;
m^{1}_{\; 2}\; {\bar z}\;\; ;
\nonumber\\
\xi^{0}_{D} &=& \alpha\; x^{0}\quad ,\quad \xi^{z}_{D}\;\; =\;\; \alpha\;
z\quad ,\quad \xi^{{\bar z}}_{D}\;\; =\;\; \alpha\; {\bar z}
\;\; ;\nonumber\\
\xi^{0}_{S} &=& x^{0}\; [b^{1}\; (z\; +\;{\bar z})\; -\; i\; b^{2}\; (z\;
-\;{\bar z})]\;\; ,\nonumber\\
\xi^{z}_{S}&=& b^{1}\; (z^{2}\; -\; x_{0}^{2})\; -\; i\;
b^{2}\; (z^{2}\; +\; x_{0}^{2})\;\; ,\nonumber\\
\xi^{{\bar z}}_{S}&=& b^{1}\; ({\bar z}^{2}\; -\; x_{0}^{2})\; +\;
i\; b^{2}\; ({\bar z}^{2}\; +\; x_{0}^{2})\;\; . \label{28} \ea The
formulation is again on surfaces at equal values of the radial
coordinate $x^{0}$. We generically define the charges at
$x^{0}=\epsilon$ as \ba Q\; =\;\oint_{0} dz \oint_{0} d{\bar
z}\;\sqrt{g}\; J^{0}\;\; , \label{794} \ea where the integrations
are counterclockwise around $z=0$ and ${\bar z}=0$. The Noether
currents and the conjugate momentum are formally given by (\ref{11},
\ref{11a}) and (\ref{8}) (supplemented by (\ref{16})), but we now
make use of the metric (\ref{27}) and the expression (\ref{28}) for
the Killing vectors. Notice that (\ref{4}, \ref{23}) now reduce to
\ba \Delta_{\pm}\; =\; 1\; \pm\; \sqrt{1\; +\; l^{2}m^{2}}\;\; ,
\label{70} \ea with \ba -1\; <\; l^{2}m^{2}\;<\; 0\;\; ,\label{71}
\ea so that \ba 0\; <\; \Delta_{-}\; <\; 1\qquad ,\qquad 1\; <\;
\Delta_{+}\; <\; 2\;\; .\label{72}\ea In particular we have the
following useful relation \ba \Delta_{+}\; +\; \Delta_{-}\; =\;
2\;\; . \label{33} \ea

The calculations are analogous to those performed in the previous
section, and we write here the results. If we consider the choice
(\ref{16}) then the asymptotic behaviors of $\Phi$ and $\Pi$ are
similar to those in (\ref{14}, \ref{18}), namely \ba \Phi(\epsilon,
z,{\bar z})\;\sim\;\epsilon^{\Delta_{-}}\;\Phi_{0}(z,{\bar z})\quad
,\quad \Pi(\epsilon, z,{\bar z})\;\sim\;\epsilon^{-\Delta_{-}-1}\;
l\;\Pi_{0}(z,{\bar z})\;\; . \label{41} \ea We find the following
charges in the asymptotic limit $\epsilon\rightarrow 0$ (here we
define ${\tilde P}_{i}\equiv \lim_{\epsilon\rightarrow 0}P_{i}$ and
so on) \ba {\tilde P}_{1} &=& -i\; (G_{T}\; +\; {\bar G}_{T})\quad
,\quad {\tilde P}_{2}\;\; =\;\;  G_{T}\; -\; {\bar G}_{T}\;\;
,\nonumber\\
{\tilde M} &=& -\; (G_{DR}\; -\; {\bar G}_{DR})\quad ,\quad
{\tilde D}\;\; =\;\; -i\; (G_{DR}\; +\; {\bar G}_{DR})\;\; ,\nonumber\\
{\tilde K}^{1} &=& -i\; (G_{S}\; +\; {\bar G}_{S})\quad ,\quad
{\tilde K}^{2}\;\; =\;\; - \; (G_{S}\; -\; {\bar G}_{S})\;\; ,
\label{30} \ea where (from now on we define
$\partial\equiv\frac{\partial}{\partial z}$ and
${\bar\partial}\equiv\frac{\partial}{\partial {\bar z}}$) \ba G_{T}
&=& -i\oint_{0} dz \oint_{0} d{\bar z}\;
\Pi_{0}\;\partial\Phi_{0}\;\;
, \nonumber\\
G_{DR} &=& -i\oint_{0} dz \oint_{0} d{\bar z}\;
\Pi_{0}\left(z\;\partial\Phi_{0}\; +\;\frac{1}{2}\;
\Delta_{-}\;\Phi_{0}\right)\; ,\nonumber\\
G_{S} &=& -i\oint_{0} dz \oint_{0} d{\bar z}\; \Pi_{0}\;
z\;(z\;\partial\Phi_{0}\; +\;\Delta_{-}\;\Phi_{0})\;\; ,\label{77}
\ea and \ba {\bar G}_{T} &=& -i\oint_{0} d{\bar z} \oint_{0} dz\;
\Pi_{0}\;{\bar\partial}\Phi_{0}\;\; ,
\nonumber\\
{\bar G}_{DR} &=& -i\oint_{0} d{\bar z} \oint_{0} dz\;
\Pi_{0}\left({\bar z}\;{\bar \partial}\Phi_{0}\; +\;\frac{1}{2}\;
\Delta_{-}\;\Phi_{0}\right)\; ,\nonumber\\
{\bar G}_{S} &=& -i\oint_{0} d{\bar z} \oint_{0} dz\; \Pi_{0}\;
{\bar z}\;({\bar z}\;{\bar \partial}\Phi_{0}\;
+\;\Delta_{-}\;\Phi_{0})\;\; .\label{78} \ea Here the integration is
performed counterclockwise over circles around $z=0$ and ${\bar
z}=0$. We will show later that $G_{T}$ and ${\bar G}_{T}$ generate
translations at the boundary, whereas $G_{DR}$ and ${\bar G}_{DR}$
correspond to both dilations and rotations. In addition, $G_{S}$ and
${\bar G}_{S}$ generate special conformal transformations.

Now, integrating by parts and using (\ref{33}) we get\footnote{In
these and the following calculations we assume the product
$\Pi_{0}(z,{\bar z})\Phi_{0}(z,{\bar z})$ to admit a Laurent
expansion in $z$ and ${\bar z}$ (even when $\Pi_{0}$ and $\Phi_{0}$
are not separately required to satisfy this property), so in
particular we have $\oint_{0} dz \; \partial
(z^{n}\Pi_{0}\Phi_{0})\; =\; \oint_{0} d{\bar z}\; {\bar \partial}
({\bar z}^{n}\Pi_{0}\Phi_{0})\; =\; 0$ $(n\in{\bf Z})$ (notice that
since we are considering contour integrals then $z$ and ${\bar z}$
can be treated as independent variables). The explicit series
expansions for $\Pi_{0}$ and $\Phi_{0}$ which satisfy these
requirements will be introduced later, after a proper motivation.}
 \ba G_{T} &=& -\frac{i}{2}\;
\oint_{0} dz \oint_{0} d{\bar z}\;
(\Delta_{+}\;\partial\Phi_{0}\;\Pi_{0}\; -\;
\Delta_{-}\;\Phi_{0}\;\partial\Pi_{0})\;\; ,\nonumber\\
G_{DR} &=& -\frac{i}{2}\; \oint_{0} dz \; z \oint_{0} d{\bar z}\;
(\Delta_{+}\;\partial\Phi_{0}\;\Pi_{0}\; -\;
\Delta_{-}\;\Phi_{0}\;\partial\Pi_{0})\;\; ,\nonumber\\
G_{S} &=& -\frac{i}{2}\; \oint_{0} dz \; z^{2}\oint_{0} d{\bar z}\;
(\Delta_{+}\;\partial\Phi_{0}\;\Pi_{0} -\;
\Delta_{-}\;\Phi_{0}\;\partial\Pi_{0})\;\; , \label{34} \ea and \ba
{\bar G}_{T} &=& -\frac{i}{2}\; \oint_{0} d{\bar z} \oint_{0} dz\;
(\Delta_{+}\;{\bar\partial}\Phi_{0}\;\Pi_{0}\; -\;
\Delta_{-}\;\Phi_{0}\;{\bar\partial}\Pi_{0})\;\; ,\nonumber\\
{\bar G}_{DR} &=& -\frac{i}{2}\; \oint_{0} d{\bar z}\;{\bar z}
\oint_{0} dz\; (\Delta_{+}\;{\bar\partial}\Phi_{0}\;\Pi_{0}\; -\;
\Delta_{-}\;\Phi_{0}\;{\bar\partial}\Pi_{0})\;\; ,\nonumber\\
{\bar G}_{S} &=& -\frac{i}{2}\; \oint_{0} d{\bar z}\; {\bar z}^{2}
\oint_{0} dz\; (\Delta_{+}\; {\bar\partial}\Phi_{0}\;\Pi_{0}\; -\;
\Delta_{-}\;\Phi_{0}\;{\bar\partial}\Pi_{0})\;\; . \label{35} \ea We
may write \ba G_{T} &=& \frac{1}{2\pi i}\; \oint_{0} dz\; T(z)\;\;\;
,\;\;\; G_{DR}\;=\; \frac{1}{2\pi i}\; \oint_{0} dz\; z\; T(z)\;\;\;
,\;\;\;
G_{S}\;=\; \frac{1}{2\pi i}\; \oint_{0} dz\; z^{2}\; T(z),\nonumber\\
{\bar G}_{T} &=& \frac{1}{2\pi i}\; \oint_{0} d{\bar z}\; {\bar
T}({\bar z})\;\;\; ,\;\;\; {\bar G}_{DR}\;=\; \frac{1}{2\pi i}\;
\oint_{0} d{\bar z}\; {\bar z}\; {\bar T}({\bar z})\;\;\; ,\;\;\;
{\bar G}_{S}\;=\; \frac{1}{2\pi i}\; \oint_{0} d{\bar z}\; {\bar
z}^{2}\; {\bar T}({\bar z}), \nonumber \ea where \ba T(z) &=& \pi
\oint_{0} d{\bar z}\; (\Delta_{+}\;\partial\Phi_{0}\;\Pi_{0} -\;
\Delta_{-}\;\Phi_{0}\;\partial\Pi_{0})\;\; ,\nonumber\\
{\bar T}({\bar z}) &=& \pi \oint_{0} dz\;
(\Delta_{+}\;{\bar\partial}\Phi_{0}\;\Pi_{0} -\;
\Delta_{-}\;\Phi_{0}\;{\bar\partial}\Pi_{0}) \;\; . \nonumber \ea
Now we Laurent expand \ba T(z)\; =\; \sum_{n\in{\bf Z}}L_{n}\;
z^{-n-2}\qquad ,\qquad {\bar T}({\bar z})\; =\; \sum_{n\in{\bf
Z}}{\bar L}_{n}\;{\bar z}^{-n-2} \;\; ,\label{534}\ea where \ba
L_{n} &=& -\frac{i}{2}\; \oint_{0} dz\; z^{n+1} \oint_{0} d{\bar
z}\; (\Delta_{+}\;\partial\Phi_{0}\;\Pi_{0}\; -\;
\Delta_{-}\;\Phi_{0}\;\partial\Pi_{0})\;\; , \label{37} \ea and \ba
{\bar L}_{n}\; =\; -\frac{i}{2}\; \oint_{0} d{\bar z}\;{\bar
z}^{n+1} \oint_{0} dz\;(\Delta_{+}\;
{\bar\partial}\Phi_{0}\;\Pi_{0}\; -\;
\Delta_{-}\;\Phi_{0}\;{\bar\partial}\Pi_{0})\;\; . \label{38} \ea In
particular, notice from (\ref{34}, \ref{35}) that \ba G_{T}&=&
L_{-1}\quad ,\quad G_{DR}\;\; =\;\; L_{0}\quad ,\quad
G_{S}\;\; =\;\; L_{1}\;\; ,\nonumber\\
{\bar G}_{T}&=& {\bar L}_{-1}\quad ,\quad {\bar G}_{DR}\;\; =\;\;
{\bar L}_{0}\quad ,\quad {\bar G}_{S}\;\; =\;\; {\bar L}_{1}\;\; .
\label{36} \ea

As in the previous section, we are considering here a formulation
where $x^{0}$ plays a special role and we step on surfaces at equal
values of it. By performing the limit $\epsilon\rightarrow 0$ we
have positioned on the boundary at $x^{0}=0$, where we consider the
Poisson brackets \ba {\{}\Pi_{0}(z,{\bar z})\; , \Phi_{0}(w,{\bar
w}){\}}_{P.B.} \;=\; \delta (z-w){\bar\delta} ({\bar z}-{\bar
w})\;\; ,\label{39}\ea and \ba {\{}\Pi_{0}(z,{\bar z})\; ,
\Pi_{0}(w,{\bar w}){\}}_{P.B.}\; =\;{\{}\Phi_{0}(z,{\bar z})\; ,
\Phi_{0}(w,{\bar w}){\}}_{P.B.}\; =\; 0\;\; . \label{44}\ea Here the
$\delta$-function satisfies (in the particular case where the
integration contours are circles around the origin, as considered
here) \ba f(w)&=&\oint_{0\; ,\; |z|=|w|} dz \;f(z)\;\delta
(z-w)\quad ,\nonumber\\ -\partial f(w)&=& \oint_{0\; ,\; |z|=|w|} dz
\;f(z)\;\partial_{z}\delta (z-w)\quad ,\nonumber\ea where $f$ is a
generic function, and the integration is over a circle satisfying
$|z|=|w|$. So the Poisson brackets are computed at equal values of
the distance to the origin of the complex plane, in a sense that we
illustrate with the following example (see (\ref{37})) \ba
&&{\{}L_{n}\; ,\; \Phi_{0}(w,{\bar w}){\}}_{P.B.} =
-\;\frac{i}{2}\;\oint_{0\; ,\; |z|=|w|} dz \;z^{n+1}\oint_{0\; ,\;
|{\bar
z}|=|{\bar w}|} d{\bar z}\nonumber\\
&&\qquad\quad\;\times\;\; [\Delta_{+}\;\partial\Phi_{0}(z,{\bar
z})\;\delta (z-w)\;{\bar\delta}({\bar z}-{\bar
w})\;-\;\Delta_{-}\;\Phi_{0}(z,{\bar z})\;\partial_{z}\delta
(z-w)\;{\bar\delta}({\bar z}-{\bar w})].\nonumber\\ \label{786}\ea
Here we are integrating over counterclockwise circles around the
origin of the complex plane and satisfying $|z|=|w|$ and $|{\bar
z}|=|{\bar w}|$. It is this property of the asymptotic Poisson
brackets, of being computed at equal values of the distance to the
origin of the complex plane, that will give them a meaning from the
point of view of the boundary two dimensional CFT.

Now using Eqs.(\ref{33}, \ref{37}, \ref{38}, \ref{39}, \ref{44}) we
get \ba i{\{}L_{n}\; ,\; \Phi_{0}(z,{\bar z}){\}}_{P.B.}&=&
\frac{\Delta_{-}}{2}\;(n+1)\; z^{n}\;\Phi_{0}(z,{\bar z})\; +
\;z^{n+1}\;\partial\Phi_{0}(z,{\bar z})\;\; ,\nonumber\\i{\{}{\bar
L}_{n}\; ,\; \Phi_{0}(z,{\bar z}){\}}_{P.B.}&=&
\frac{\Delta_{-}}{2}\;(n+1)\; {\bar z}^{n}\;\Phi_{0}(z,{\bar z})\; +
\;{\bar z}^{n+1}\;{\bar\partial}\Phi_{0}(z,{\bar z})\;\;
,\label{55}\ea and \ba i{\{}L_{n}\; ,\; \Pi_{0}(z,{\bar
z}){\}}_{P.B.}&=& \frac{\Delta_{+}}{2}\;(n+1)\;
z^{n}\;\Pi_{0}(z,{\bar z})\; + \;z^{n+1}\;\partial\Pi_{0}(z,{\bar
z})\;\; ,\nonumber\\i{\{}{\bar L}_{n}\; ,\; \Pi_{0}(z,{\bar
z}){\}}_{P.B.}&=& \frac{\Delta_{+}}{2}\;(n+1)\; {\bar
z}^{n}\;\Pi_{0}(z,{\bar z})\; + \;{\bar
z}^{n+1}\;{\bar\partial}\Pi_{0}(z,{\bar z})\;\; .\label{56}\ea These
expressions suggest the identification of the coefficients $L_{n}$
and ${\bar L}_{n}$ with generators of the local conformal group on
the boundary of $AdS_{3}$, and of the asymptotic values of the bulk
scalar field and the conjugate momentum, namely $\Phi_{0}$ and
$\Pi_{0}$, respectively, with conformal fields with the following
weights \ba \Phi_{0}&\longrightarrow & h_{\Phi_{0}}\; =\;
{\bar h}_{\Phi_{0}}\; =\; \frac{\Delta_{-}}{2}\quad ,\nonumber\\
\Pi_{0}&\longrightarrow & h_{\Pi_{0}}\; =\; {\bar h}_{\Pi_{0}}\; =\;
\frac{\Delta_{+}}{2}\quad . \label{51}\ea In particular, these
results are consistent with those in the previous section (see
(\ref{75}, \ref{76})) where the global conformal group in $d$
dimensions was found and we computed the scaling dimensions (i.e.
$h_{\Phi_{0}}+{\bar h}_{\Phi_{0}}$ and $h_{\Pi_{0}}+{\bar
h}_{\Pi_{0}}$ in the particular case of $d=2$) of $\Phi_{0}$ and
$\Pi_{0}$. This can be considered as a consistency check. Notice,
also, that both fields $\Phi_{0}$ and $\Pi_{0}$ have spin zero
($h_{\Phi_{0}}-{\bar h}_{\Phi_{0}}= h_{\Pi_{0}}-{\bar h}_{\Pi_{0}}=
0$). In addition, from (\ref{36}) and (\ref{55}, \ref{56}) we see
that $(G_{T}, G_{DR}, G_{S})$ and $({\bar G}_{T}, {\bar G}_{DR},
{\bar G}_{S})$ correspond to the global conformal group and that, as
anticipated, $G_{T}$ and ${\bar G}_{T}$ generate translations at the
boundary, whereas $G_{DR}$ and ${\bar G}_{DR}$ generate both
dilations and rotations, and $G_{S}$, as well as ${\bar G}_{S}$,
correspond to special conformal transformations.

We also find the expected results\footnote{This is more easily
computed by integrating by parts (see footnote 11) e.g. in
(\ref{37}) and using (\ref{33}) in order to write \ba iL_{n}\;=\;
A_{n}\; +\; \frac{\Delta_{-}}{2}\; (n+1)\; B_{n}\;\; ,\nonumber\ea
where \ba A_{n}\; =\; \oint_{0} dz\; z^{n+1} \oint_{0} d{\bar z}\;
\partial\Phi_{0}\;\Pi_{0}\quad ,\quad B_{n}\; =\; \oint_{0} dz
\; z^{n} \oint_{0} d{\bar z}\; \Phi_{0}\;\Pi_{0}\;\; .
 \nonumber\ea Then by computing
\ba {\{}A_{n}\; ,\; A_{m}{\}}_{P.B.}\; =\; (n-m)\; A_{n+m}\quad
,\quad{\{}A_{n}\; ,\; B_{m}{\}}_{P.B.}\; =\; -m\; B_{n+m}\quad
,\quad{\{}B_{n}\; ,\; B_{m}{\}}_{P.B.}\; =\; 0\;\; ,\nonumber\ea
Eq.(\ref{47}) follows. In a similar way (\ref{48}) and (\ref{49})
can also be obtained. With illustrative purposes, notice that the
integration contours are taken as, e.g., \ba &&{\{}A_{n}\; ,\;
A_{m}{\}}_{P.B.} \;=\; \oint_{0} dw \; w^{m+1}\oint_{0} d{\bar
w}\;\oint_{0,|z|=|w|} dz\; z^{n+1}\oint_{0,|{\bar z}|=|{\bar
w}|}d{\bar z}\nonumber\\ && \qquad\times\;
[-\partial\Phi_{0}(w,{\bar w})\;\Pi_{0}(z,{\bar
z})\;\partial_{z}\delta (z-w)\;{\bar\delta}({\bar z}-{\bar w})\; +\;
\partial\Phi_{0}(z,{\bar z})\;\Pi_{0}(w,{\bar
w})\;\partial_{w}\delta (z-w)\;{\bar\delta} ({\bar z}-{\bar w})].
\nonumber\ea } \ba i{\{}L_{n}\; ,\; L_{m}{\}}_{P.B.}\; =\; (n-m)\;
L_{n+m}\;\; ,\label{47}\ea \ba i{\{}{\bar L}_{n}\; ,\; {\bar
L}_{m}{\}}_{P.B.}\; =\; (n-m)\; {\bar L}_{n+m}\;\; ,\label{48}\ea
and \ba {\{}L_{n}\; ,\; {\bar L}_{m}{\}}_{P.B.}\; =\; 0\;\;
.\label{49}\ea

In this way, after performing the asymptotic limit we have obtained
two copies of the Virasoro algebra, with vanishing central charges
at the classical level. These results, as well as (\ref{55}) and
(\ref{56}), can be considered as consistency checks on the
formalism. As a further additional check, Eqs.(\ref{47}, \ref{48},
\ref{49}) can also be obtained using mode expansions. We
write\footnote{Using (\ref{33}) we see that these mode expansions
meet the requirements in footnote 11. There is the formal aspect of
the well definiteness of the product of two Laurent expansions, but,
since this will not affect the physics, then we will not consider it
here.} \ba \Phi_{0}(z,{\bar z})&=&\frac{1}{2\pi}\;\sum_{n,m\in {\bf
Z}}z^{-n-\frac{\Delta_{-}}{2}}\;{\bar
z}^{-m-\frac{\Delta_{-}}{2}}\;\Phi_{n,m}\quad , \nonumber\\
\Pi_{0}(z,{\bar z})&=&\frac{1}{2\pi}\;\sum_{n,m\in {\bf
Z}}z^{-n-\frac{\Delta_{+}}{2}}\;{\bar
z}^{-m-\frac{\Delta_{+}}{2}}\;\Pi_{n,m}\quad , \label{50}\ea where
\ba \Phi_{n,m}&=& -\frac{1}{2\pi}\;\oint_{0} dz\;
z^{n+\frac{\Delta_{-}}{2}-1} \oint_{0} d{\bar z}\; {\bar
z}^{m+\frac{\Delta_{-}}{2}-1}\; \Phi_{0}(z,{\bar z})\quad ,\nonumber\\
\Pi_{n,m}&=& -\frac{1}{2\pi}\;\oint_{0} dz\;
z^{n+\frac{\Delta_{+}}{2}-1} \oint_{0} d{\bar z}\; {\bar
z}^{m+\frac{\Delta_{+}}{2}-1}\; \Pi_{0}(z,{\bar z})\quad .\label{52}
\ea The extra powers of $\frac{\Delta_{-}}{2}$ and
$\frac{\Delta_{+}}{2}$ appearing in (\ref{50}) are motivated by
(\ref{51}).

From (\ref{33}), (\ref{39}, \ref{44}) and (\ref{52}) we find \ba
{\{}\Pi_{n,m}\; ,\;\Phi_{r,s}{\}}_{P.B.}\; =\;
-\;\delta_{n+r,0}\;{\bar\delta}_{m+s,0}\;\; ,\label{53}\ea and \ba
{\{}\Pi_{n,m}\; ,\;\Pi_{r,s}{\}}_{P.B.}\; =\;{\{}\Phi_{n,m}\;
,\;\Phi_{r,s}{\}}_{P.B.}\; =\; 0\;\; . \label{533}\ea On the other
hand, plugging (\ref{50}) into (\ref{37}, \ref{38}) and using
(\ref{33}) we get \ba L_{n} & =& i\sum_{r,s\in {\bf
Z}}\left(n\;\frac{\Delta_{-}}{2}\; -\; r\right)\Phi_{r,s}\;
\Pi_{n-r,-s}\;\;\; ,\nonumber\\ {\bar L}_{n} &=& i\sum_{r,s\in {\bf
Z}}\left(n\;\frac{\Delta_{-}}{2}\; -\; s\right)
\Phi_{r,s}\;\Pi_{-r,n-s}\;\; .\label{54}\ea Now, using (\ref{53},
\ref{533}, \ref{54}) and performing some straightforward algebra we
reproduce Eqs.(\ref{47}, \ref{48}, \ref{49}) again, as expected.

We also find the following results \ba i{\{}L_{n}\; ,\;
\Phi_{r,s}{\}}_{P.B.} &=& \left[n\left(\frac{\Delta_{-}}{2}\; -\;
1\right)-r\right]\; \Phi_{n+r,s}\;\; ,\nonumber\\ i{\{}{\bar
L}_{n}\; ,\; \Phi_{r,s}{\}}_{P.B.} &=&
\left[n\left(\frac{\Delta_{-}}{2}\; -\; 1\right)-s\right]\;
\Phi_{r,n+s}\;\; , \label{60}\ea and \ba i{\{}L_{n}\; ,\;
\Pi_{r,s}{\}}_{P.B.} &=& \left[n\left(\frac{\Delta_{+}}{2}\; -\;
1\right)-r\right]\; \Pi_{n+r,s}\;\; ,\nonumber\\ i{\{}{\bar L}_{n}\;
,\; \Pi_{r,s}{\}}_{P.B.}&=& \left[n\left(\frac{\Delta_{+}}{2}\; -\;
1\right)-s\right]\; \Pi_{r,n+s}\;\; , \label{61}\ea which are
consistent with (\ref{55}, \ref{56}).

On the basis of the results above, we interpret the coefficients
$L_{n},\; {\bar L}_{n}$ in (\ref{54}) to act as the Virasoro
generators of the boundary CFT. Notice that Hermitian conjugation
\ba \Phi^{\dag}_{n,m}\; =\;\Phi_{-n,-m}\qquad ,\qquad
\Pi^{\dag}_{n,m}\; =\;\Pi_{-n,-m}\;\; ,\label{63}\ea gives in
(\ref{54}) the expected properties \ba L^{\dag}_{n}\; =\; L_{-n}
\qquad ,\qquad {\bar L}^{\dag}_{n}\; =\; {\bar L}_{-n}\;\;
.\label{64}\ea

These results motivate us to consider aspects of the quantization of
the boundary CFT and attempt to compute the quantum central charge.
From now on, $\Phi_{0}$ and $\Pi_{0}$ will be treated as conformal
fields of the boundary CFT having weights as given by (\ref{51}),
their role as the asymptotic values of fields living in the bulk
being put aside. Notice that Eqs.(\ref{55}, \ref{56}) and (\ref{60},
\ref{61}) suggest to consider the following $AdS_{3}/CFT_{2}$
prescription \ba {\{}\;\; ,\;\;{\}}_{P.B.}\;\;\longrightarrow\;\;
-i\;[\;\;,\;\;]\;\; ,\label{80}\ea where $[\;\;,\;\;]$ are
commutators on the boundary CFT, consistent with the radial
quantization procedure.\footnote{The expression `radial
quantization' is not to be confused with the radial coordinate of
$AdS_{3}$, since it refers to the standard quantization procedure in
two dimensional CFTs, where operators within correlation functions
are radially ordered.} In particular, this gives in (\ref{53},
\ref{533}) \ba [\Pi_{n,m}\; ,\; \Phi_{r,s}]\; =\; -\; i\;
\delta_{n+r,0}\;{\bar\delta}_{m+s,0}\;\; ,\label{62}\ea and \ba
[\Pi_{n,m}\; ,\;\Pi_{r,s}]\; =\;[\Phi_{n,m}\; ,\;\Phi_{r,s}]\; =\;
0\;\; . \label{622}\ea

We point out that the reason why (\ref{80}) makes sense is the fact
that the asymptotic Poisson brackets are computed at equal values of
the distance to the origin of the complex plane, as discussed above
and illustrated e.g. by (\ref{786}). It is this property, which is
inherited from the definition of the charges using contour integrals
(see (\ref{794})), that causes the asymptotic Poisson brackets to
have a meaning from the point of view of the two dimensional
boundary CFT, and is what allows us to go here one step further than
in the generic $AdS_{d+1}$ case analyzed in the previous section.

From now on, we will simultaneously consider the following two
alternative sets of definitions \ba && b_{n,m}\;\equiv\;
i\Phi_{n,m}\qquad ,\qquad c_{n,m}\;\equiv\;\Pi_{n,m}\;\; ,\nonumber\\
&&\qquad\qquad\qquad\lambda\;\equiv\;\frac{\Delta_{-}}{2}\;\;
,\label{65}\ea or else \ba && b_{n,m}\;\equiv\;
-i\Pi_{n,m}\qquad ,\qquad c_{n,m}\;\equiv\;\Phi_{n,m}\;\; ,\nonumber\\
&&\;\;\;\qquad\qquad\qquad\lambda\;\equiv\;\frac{\Delta_{+}}{2}\;\;
.\label{66}\ea Choosing any of the definitions (\ref{65}) or
(\ref{66}), and using (\ref{33}), we see that Eqs.(\ref{54},
\ref{63}, \ref{62}) can be written in the following way \ba L_{n} &
=& \sum_{r,s\in {\bf Z}}\left(n\;\lambda\; -\; r\right)b_{r,s}\;
c_{n-r,-s}\;\;\; ,\nonumber\\ {\bar L}_{n} &=& \sum_{r,s\in {\bf
Z}}\left(n\;\lambda\; -\; s\right) b_{r,s}\;c_{-r,n-s}\;\;
,\label{67}\ea \ba b^{\dag}_{n,m}\; =\;-b_{-n,-m}\qquad ,\qquad
c^{\dag}_{n,m}\; =\;c_{-n,-m}\;\; ,\label{68}\ea \ba [c_{n,m}\; ,\;
b_{r,s}]\; =\; \delta_{n+r,0}\;{\bar\delta}_{m+s,0}\; .\label{69}\ea
Exception made of the fact that the fields are neither purely
holomorphic nor antiholomorphic, this is strikingly similar to a
generalized ghost system.\footnote{The existence of the analogous of
a ghost current will not be discussed here.}

We would like to compute the quantum central charge in the two
copies of the Virasoro algebra \ba [L_{n}\; ,\; L_{m}] \;=\; (n-m)\;
L_{n+m}\; +\; \frac{c}{12}\; n\; (n^{2}\; -\; 1)\;\delta_{n+m,0}\;\;
,\label{81}\ea and \ba [{\bar L}_{n}\; ,\; {\bar L}_{m}]\; =\;
(n-m)\; {\bar L}_{n+m}\;+\; \frac{{\bar c}}{12}\; n\; (n^{2}\; -\;
1)\;{\bar\delta}_{n+m,0}\;\; .\label{82}\ea We define the following
ground state \ba b_{0,0}|0\rangle\; =\; 0\qquad\quad ,\quad\qquad
b_{n,m}|0\rangle\; =\; c_{n,m}|0\rangle\; =\; 0\;\;\; \quad (n>0\;\;
\hbox{or} \;\; m>0)\;\; ,\label{83}\ea which is not $SL_{2}$
invariant. Here we have grouped $b_{0,0}$ with the lowering
operators and $c_{0,0}$ with the raising ones. Notice that, in fact,
this gives two possible ground states, corresponding to the choices
(\ref{65}, \ref{66}). This should correspond to the two possible
quantizations in the bulk found in \cite{freedman}\cite{freedman2}.
The fact that we find two possible quantizations, as expected, could
be considered as an additional consistency check on our formalism.

We introduce a normal ordering in (\ref{67}) where lowering
operators are placed to the right. We find \ba L_{n}|0\rangle\; =\;
{\bar L}_{n}|0\rangle\; =\; 0\qquad (n>0)\;\; .\label{84}\ea The
result \ba L_{1}L_{-1}|0\rangle\; =\; {\bar L}_{1}{\bar
L}_{-1}|0\rangle\; =\; -\lambda (1-\lambda)|0\rangle\;\;
,\nonumber\ea gives the following normal ordering constant for the
coefficients $L_{0}$ and ${\bar L}_{0}$ \ba L_{0}|0\rangle\; =\;
{\bar L}_{0}|0\rangle\; =\; -\frac{1}{2}\;\lambda\;
(1-\lambda)|0\rangle\;\; .\label{85}\ea Using any of the choices
(\ref{65}) or (\ref{66}), together with (\ref{70}), we obtain \ba
L_{0}|0\rangle\; =\; {\bar L}_{0}|0\rangle\; =\;
\frac{l^{2}m^{2}}{8}|0\rangle\;\; .\label{86}\ea Notice from
(\ref{71}) that this constant is in the range \ba
-\frac{1}{8}\;<\;\frac{l^{2}m^{2}}{8}\;<0\;\;.\label{87}\ea

Now computing \ba L_{2}L_{-2}|0\rangle\; =\; {\bar L}_{2}{\bar
L}_{-2}|0\rangle\; =\; [(2\lambda -1)^{2}-\; 4\lambda
(1-\lambda)]|0\rangle\;\; ,\nonumber\ea we get the central charges
\ba c\; =\; {\bar c}\; =\; 3(2\lambda -1)^{2}-\; 1\; .\label{73}\ea
Using any of the choices (\ref{65}) or (\ref{66}), together with
(\ref{70}), we find \ba c\; =\; {\bar c}\; =\; 2\; +\;
3l^{2}m^{2}\;\; .\label{88}\ea We also notice from (\ref{71}) that
the central charges are in the range \ba -1\;<\; c\;<\; 2\qquad
,\qquad -1\;<\; {\bar c}\;<\; 2\;\; .\label{89}\ea

As seen in (\ref{86}, \ref{88}) both the normal ordering constant
and quantum central charge depend on the mass of the bulk field and
on the AdS radius, in an analogous way as the conformal dimensions
do.

A striking result, which is obtained from the facts that the
boundary CFT is similar to a generalized ghost system (see
(\ref{67}, \ref{68}, \ref{69})), and from the explicit dependence of
the central charge with the mass (see (\ref{88})) is that the mass
of the bulk scalar field seems to play the role of a background
charge for the boundary CFT. In order to shed some light into this
susprising result, it would be interesting to perform a bosonization
program on the boundary CFT. However, this issue will no longer be
considered here.

Notice also from (\ref{88}) that the requirement for the central
charge to be positive gives \ba l^{2}m^{2}\;
>\; -\frac{2}{3}\;\; .\label{90}\ea This is more restrictive than
the Breitenlohner-Freedman bound (see (\ref{1771})). This new
interesting result seems to be detected only from the boundary point
of view, and the way it could be red from the bulk point of view
remains to be investigated.

We also point out again that we have found two possible choices for
the ground state, corresponding to (\ref{65}, \ref{66}), together
with (\ref{83}), which should in turn correspond to the two possible
quantizations in the bulk found in \cite{freedman}\cite{freedman2},
and that this result could be considered as a last consistency check
on the formalism that we have developed here.

It would be interesting to investigate the generalization of the
present formalism to the case of interacting scalar field in the
bulk. For instance, we could consider a polynomial interaction \ba
I\;=\;\int d^{3} x\; \sqrt{g}\left[\frac{1}{2}\left(
g^{\mu\nu}\partial_{\mu}\Phi\;\partial_{\nu}\Phi \;+\;
m^{2}\Phi^{2}\right)+\sum_{n\geq
3}\frac{\alpha_{n}}{n!}\;\Phi^{n}\right]\;\; .\nonumber\ea Notice
that, exception made of the asymptotic behavior (\ref{7}), we have
not made use of the explicit form of the solution to the equation of
motion in this formalism. This property is expected to simplify the
calculations. On the other hand, we could consider a perturbative
approach where the solution to the free scalar field case is
inserted into modified `conserved' charges, thus possibly giving
rise to modified theories at the boundary.

Another interesting issue is the inclusion of gravity in the present
formalism. We would like to analyze the case of asymptotically
anti-de Sitter spaces. At a more speculative level, there is the
interesting possibility that the results we have computed here
correspond to a certain limit of a gravity theory. This is suggested
by the facts that we have found a full Virasoro symmetry algebra in
the asymptotic limit of $AdS_{3}$, and further, that the
corresponding energy-momentum tensor (see e.g.(\ref{534})) is in
principle expected to correspond to a massless spin 2 field in the
bulk. It is possible that such investigations could allow us to
establish a connection among the formalism developed here and the
results in \cite{bh}, and, in addition, to investigate possible
applications in black hole physics.

Finally, it also would be interesting to consider possible
extensions of our formalism to higher dimensional Anti-de Sitter
spaces, and to values of $l^{2}m^{2}$ outside the range (\ref{4}).
\\
\vspace*{0.5cm}

{\large {\bf Acknowledgments}:} It is a pleasure to thank W. Baron,
C. Cardona, G. Giribet, S. Iguri, L. Nicol\'as and C. N\'u\~nez for
very useful comments and discussions, and M. Schvellinger for
encouragement. This work was supported by CONICET.


\begin{thebibliography}{99}
\bibitem{maldacena} J. M. Maldacena, ``The Large N Limit of Superconformal
Field Theories and Supergravity'', Adv. Theor. Math. Phys. {\bf 2}
(1998) 231; Int. J. Theor. Phys. {\bf 38} (1999) 1113.
\bibitem{witten} E. Witten, ``Anti De Sitter Space and Holography'',
Adv. Theor. Math. Phys. {\bf 2} (1998) 253.
\bibitem{gubser} S. S. Gubser, I. R. Klebanov and A. M. Polyakov,
``Gauge Theory Correlators from Non-Critical String Theory'', Phys.
Lett. {\bf B428} (1998) 105.
\bibitem{freedman} P. Breitenlohner and D. Z. Freedman, ``Stability in
Gauged Extended Supergravity'', Ann. Phys. {\bf 144} (1982) 249.
\bibitem{freedman2} P. Breitenlohner and D. Z. Freedman, ``Positive Energy
in Anti-de Sitter Backgrounds and Gauged Extended Supergravity'',
Phys. Lett. {\bf B115} (1982) 197.
\bibitem{mezincescu} L. Mezincescu and P. K. Townsend, ``Stability at a
Local Maximum in Higher Dimensional Anti-de Sitter Space and Applications
to Supergravity'', Ann. Phys. {\bf 160} (1985) 406.
\bibitem{freedman3} D. Z. Freedman, S. D. Mathur, A. Matusis and
L. Rastelli, ``Correlation functions in the $CFT_{d}/AdS_{d+1}$
correspondence'', Nucl. Phys. {\bf B546} (1999) 96.
\bibitem{mueck} W. M\"uck and K. S. Viswanathan, ``Conformal Field
Theory Correlators from Classical Scalar Field Theory on $AdS_{d+1}$'',
Phys. Rev. {\bf D58} (1998) 041901.
\bibitem{kw} I. R. Klebanov and E. Witten, ``AdS/CFT Correspondence
and Symmetry Breaking'', Nucl. Phys. {\bf B556} (1999) 89.
\bibitem{kraus} V. Balasubramanian, P. Kraus and A. Lawrence, ``Bulk vs.
Boundary Dynamics in Anti-de Sitter Spacetime'', Phys. Rev. {\bf
D59} (1999) 046003.
\bibitem{witten2} E. Witten, ``Multi-Trace Operators, Boundary
Conditions, And AdS/CFT Correspondence'', arXiv:hep-th/0112258.
\bibitem{berkooz} M. Berkooz, A. Sever and A. Shomer, ``Double-trace
Deformations, Boundary Conditions and Spacetime Singularities'',
JHEP {\bf 0205} (2002) 034.
\bibitem{mueck2} W. M\"uck, ``An improved correspondence formula for
AdS/CFT with multi-trace operators'', Phys. Lett. {\bf B531} (2002)
301.
\bibitem{m1} P. Minces, ``Multi-Trace Operators and the Generalized
AdS/CFT Prescription'', Phys. Rev. {\bf D68} (2003) 024027.
\bibitem{petkou} A. C. Petkou, ``Boundary multi-trace deformations
and OPEs in AdS/CFT correspondence'', JHEP {\bf 0206} (2002) 009.
\bibitem{sever} A. Sever and A. Shomer, ``A Note on Multi-trace
Deformations and AdS/CFT'', JHEP {\bf 0207} (2002) 027.
\bibitem{gubser1} S. S. Gubser and I. Mitra, ``Double-trace
operators and one-loop vacuum energy in AdS/CFT'', Phys. Rev. {\bf
D67} (2003) 064018.
\bibitem{gubser2} S. S. Gubser and I. R. Klebanov, ``A universal
result on central charges in the presence of double-trace
deformations'', Nucl. Phys. {\bf B656} (2003) 23.
\bibitem{witten5} E. Witten, ``SL(2,Z) Action On Three-Dimensional Conformal
Field Theories With Abelian Symmetry'', arXiv:hep-th/0307041.
\bibitem{satoh} Y. Satoh and J. Troost, ``On time-dependent
AdS/CFT'', JHEP {\bf 0301} (2003) 027.
\bibitem{troost} J. Troost, ``A Note on causality in the bulk and
stability on the boundary'', Phys. Lett. {\bf B578} (2004) 210.
\bibitem{nolland} D. Nolland, ``AdS/CFT boundary conditions,
multi-trace perturbations, and the c-theorem'', Phys. Lett. {\bf
B584} (2004) 192.
\bibitem{m2} P. Minces, ``Bound States in the AdS/CFT Correspondence'',
Phys. Rev. {\bf D70} (2004) 025011.
\bibitem{skenderis} I. Papadimitriou and K. Skenderis, ``AdS/CFT
correspondence and Geometry'', arXiv:hep-th/0404176.
\bibitem{zoubos} K. Zoubos, ``A Conformally Invariant Holographic
Two-Point Function on the Berger Sphere'', JHEP {\bf 0501} (2005)
031.
\bibitem{hollands} S. Hollands, A. Ishibashi and D. Marolf,
``Comparison between various notions of conserved charges in
asymptotically AdS-spacetimes'', Class. Quant. Grav. {\bf 22} (2005)
2881.
\bibitem{marolf1} D. Marolf, ``States and Boundary Terms: Subtleties
of Lorentzian AdS/CFT'', JHEP {\bf 0505} (2005) 042.
\bibitem{aharony} O. Aharony, M. Berkooz and B. Katz, ``Non-Local
Effects of Multi-Trace Deformations in the AdS/CFT Correspondence'',
JHEP {\bf 0510} (2005) 097.
\bibitem{marolf2} D. Marolf and S. Ross, ``Boundary Conditions and
Dualities: Vector Fields in AdS/CFT'', JHEP {\bf 0611} (2006) 085.
\bibitem{hartman} T. Hartman and L. Rastelli, ``Double-Trace
Deformations, Mixed Boundary Conditions and Functional Determinants
in AdS/CFT'', JHEP {\bf 0801} (2008) 019.
\bibitem{marolf3} D. Marolf and S. F. Ross, ``Reversing
Renormalization-Group Flows with AdS/CFT'', JHEP {\bf 0805} (2008)
055.
\bibitem{kiritsis} E. Kiritsis, ``Product CFTs, gravitational cloning,
massive gravitons and the space of gravitational duals'', JHEP {\bf
0611} (2006) 049.
\bibitem{banados} M. Banados, A. Schwimmer and S. Theisen,
``Remarks on Resonant Scalars in the AdS/CFT Correspondence'', JHEP
{\bf 0609} (2006) 058.
\bibitem{papadimi} I. Papadimitriou, ``Non-Supersymmetric Membrane
Flows from Fake Supergravity and Multi-Trace Deformations'', JHEP
{\bf 0702} (2007) 008.
\bibitem{diaz} D. E. Diaz and H. Dorn, ``Partition functions and
double-trace deformations in AdS/CFT'', JHEP {\bf 0705} (2007) 046.
\bibitem{papadi} I. Papadimitriou, ``Multi-Trace Deformations in
AdS/CFT: Exploring the Vacuum Structure of the Deformed CFT'', JHEP
{\bf 0705} (2007) 075.
\bibitem{diaz1} D. E. Diaz, ``Polyakov formulas for GJMS operators
from AdS/CFT'', JHEP {\bf 0807} (2008) 103.
\bibitem{kiritsis2} E. Kiritsis and V. Niarchos, ``(Multi)Matrix
Models and Interacting Clones of Liouville Gravity'', JHEP {\bf
0808} (2008) 044.
\bibitem{frolov} G. E. Arutyunov and S. A. Frolov, ``On the origin of
supergravity boundary terms in the AdS/CFT correspondence'', Nucl.
Phys. {\bf B544} (1999) 576.
\bibitem{m3} P. Minces and V. O. Rivelles, ``Scalar Field Theory in the
AdS/CFT Correspondence Revisited'', Nucl. Phys. {\bf B572} (2000)
651.
\bibitem{mr} P. Minces and V. O. Rivelles, ``Energy and the AdS/CFT
Correspondence'', JHEP {\bf 0112} (2001) 010.
\bibitem{freedman5} E. D'Hoker and D. Z. Freedman, ``Supersymmetric
Gauge Theories and the AdS/CFT Correspondence'',
arXiv:hep-th/0201253.
\bibitem{wald} A. Ishibashi and R. M. Wald, ``Dynamics in
Non-Globally-Hyperbolic Static Spacetimes III: Anti-de Sitter
Spacetime'', Class. Quant. Grav. {\bf 21} (2004) 2981.
\bibitem{andrade} T. Andrade, M. Banados and F. Rojas, ``Variational
Methods in AdS/CFT'', Phys. Rev. {\bf D75} (2007) 065013.
\bibitem{bh} J. D. Brown and M. Henneaux, ``Central Charges in the
Canonical Realization of Asymptotic Symmetries: An Example from
Three Dimensional Gravity'', Comm. Math. Phys. {\bf 104} (1986) 207.
\bibitem{henneaux} M. Henneaux, C. Martinez, R. Troncoso and J.
Zanelli, ``Black holes and asymptotics of 2+1 gravity coupled to a
scalar field'', Phys. Rev. {\bf D65} (2002) 104007.
\bibitem{henneaux2} M. Henneaux, C. Martinez, R. Troncoso and J.
Zanelli, ``Asymptotically Anti-de Sitter spacetimes and scalar
fields with a logarithmic branch'', Phys. Rev. {\bf D70} (2004)
044034.

\end{thebibliography}
\end{document}